\documentclass[aps,prd,superscriptaddress,twocolumn, nofootinbib]{revtex4-2}
\usepackage[utf8]{inputenc}
\usepackage{dsfont}
\usepackage{amsfonts}
\usepackage{amsmath}
\allowdisplaybreaks[4]        
\usepackage{amssymb}
\usepackage{euscript}     
\usepackage{braket}
\usepackage{starfont}
\usepackage{color,soul}         
\usepackage{tensor}        
\usepackage{amsthm}
\usepackage{graphicx}
\usepackage{slashed}
\usepackage{leftidx}
\usepackage{subfigure}
\usepackage{bbm}
\definecolor{outerspace}{rgb}{0.25, 0.29, 0.3}
\definecolor{scarlet}{rgb}{1.0, 0.13, 0.0}
\usepackage[header,title,page,titletoc]{appendix}  
\definecolor{princetonorange}{rgb}{1.0, 0.56, 0.0}
\definecolor{WildStrawberry}{rgb}{1.0, 0.26, 0.64}
\definecolor{rossocorsa}{rgb}{0.83, 0.0, 0.0}
\definecolor{navyblue}{rgb}{0.0, 0.0, 0.5}
\usepackage{float}
\usepackage[paper=letterpaper,margin=1in]{geometry}
\usepackage{mathtools}

\DeclareMathAlphabet{\pazocal}{OMS}{zplm}{m}{n}
\newcommand{\req}[1]{(\ref{#1})} %{Eq.\thinspace(\ref{#1})}  

\newcommand{\bea}{\begin{eqnarray}}
\newcommand{\diff}{\mathrm{d}}
\newcommand{\eea}{\end{eqnarray}}
\newcommand{\ba}{\begin{eqnarray}}
\newcommand{\ea}{\end{eqnarray}}

\newcommand{\be}{\begin{equation}}
\newcommand{\ee}{\end{equation} }
\newcommand{\beqa}{\begin{eqnarray}}
\newcommand{\eeqa}{\end{eqnarray}}
\newcommand{\beqar}{\begin{eqnarray*}}
\newcommand{\eeqar}{\end{eqnarray*}}

\renewcommand{\req}[1]{(\ref{#1})}

\newcommand{\ie}{{\it i.e.,}\ }

\makeatletter
\newcommand{\dal}{\mathop{\mathpalette\dal@\relax}}
\newcommand{\dal@}[2]{%
  \begingroup
  \sbox\z@{$\m@th#1\square$}%
  \dimen0=\fontdimen8
    \ifx#1\displaystyle\textfont\else
    \ifx#1\textstyle\textfont\else
    \ifx#1\scriptstyle\scriptfont\else
    \scriptscriptfont\fi\fi\fi3
  \makebox[\wd\z@]{%
    \hbox to \ht\z@{%
      \vrule width \dimen0
      \kern-\dimen0
      \vbox to \ht\z@{
        \hrule height \dimen0 width \ht\z@
        \vss
        \hrule height 2\dimen0
      }%
      \kern-2.5\dimen0
      \vrule width 2.5\dimen0
    }%
  }%
  \endgroup
}
\makeatother

\usepackage{hyperref}
\hypersetup{
    colorlinks,
    citecolor=navyblue,
    filecolor=navyblue,
    linkcolor=navyblue,
    urlcolor=navyblue
}

\begin{document}

\title{Regular black hole formation in four-dimensional non-polynomial gravities}
\author{Pablo Bueno}
\email{pablobueno@ub.edu}
\affiliation{Departament de F\'isica Qu\`antica i Astrof\'isica, Institut de Ci\`encies del Cosmos\\
 Universitat de Barcelona, Mart\'i i Franqu\`es 1, E-08028 Barcelona, Spain }

\author{Pablo A.~Cano}
\email{pablocano@um.es}
\affiliation{Departament de F\'isica Qu\`antica i Astrof\'isica, Institut de Ci\`encies del Cosmos\\
 Universitat de Barcelona, Mart\'i i Franqu\`es 1, E-08028 Barcelona, Spain }
 \affiliation{Departamento de F\'isica, Universidad de Murcia, Campus de Espinardo, 30100 Murcia, Spain}

\author{Robie A.~Hennigar}
\email{robie.a.hennigar@durham.ac.uk}
\affiliation{Centre for Particle Theory, Department of Mathematical Sciences, Durham University, Durham DH1 3LE, U.K.}

\author{\'Angel J.~Murcia}
\email{angelmurcia@icc.ub.edu}
\affiliation{Departament de F\'isica Qu\`antica i Astrof\'isica, Institut de Ci\`encies del Cosmos\\
 Universitat de Barcelona, Mart\'i i Franqu\`es 1, E-08028 Barcelona, Spain }

%\date{\today}
\begin{abstract}
We construct four-dimensional gravity theories that resolve the Schwarzschild singularity and enable dynamical studies of nonsingular gravitational collapse. The construction employs a class of nonpolynomial curvature invariants that produce actions with (i) second-order equations of motion in spherical symmetry and (ii) a Birkhoff theorem, ensuring uniqueness of the spherically symmetric solution. Upon spherical reduction to two dimensions, these theories map to a particular subclass of Horndeski scalar–tensor models, which we use to explicitly verify the formation of regular black holes as the byproduct of the collapse of pressureless stars and thin-shells. We also show that linear perturbations on top of maximally symmetric backgrounds are governed by second-order equations. 
\end{abstract}
\maketitle

\section{Introduction}

The regular black hole (RBH) program is the oldest approach to the problem of singularity resolution. Since as early as the 1960s, authors have written down singularity-free metrics and studied their properties~\cite{Sakharov:1966aja, 1968qtr..conf...87B, Poisson:1988wc,Dymnikova:1992ux,Hayward:2005gi}. Over the intervening years, it has been possible on kinematical considerations alone to develop quite a substantial physical picture of these objects~\cite{Carballo-Rubio:2018pmi,Carballo-Rubio:2021bpr, Carballo-Rubio:2022kad,Carballo-Rubio:2025fnc,LimaJunior:2025uyj}. However, an important question has remained:~if singularities can be resolved when the classical metric description remains sensible (even if approximately so), should not classical field theories exist which predict RBHs?

% This issue is simultaneously delicate and acute. It is delicate because one must exercise care when interpreting the results of a \textit{particular} theoretical framework. It is acute because

This problem is particularly serious as the most important open problems in the theory of RBHs---and the major criticisms of the program itself---concern \textit{dynamics}. How do RBHs form? Are they unique? Are they stable? What is the fate of their inner horizons? What is the end point of their evaporation? Do the same theories that resolve black hole singularities resolve other singularities, such as cosmological ones? Without a dynamical framework, none of these problems can be addressed. 

Despite considerable effort, this problem has persisted. This is not to say that there have been \textit{no} cases of RBHs being obtained as solutions, but rather that no general or satisfactory mechanism has been identified. For example, it is popular in the literature to consider RBHs as solutions to theories that involve general relativity (GR) coupled to matter fields, such as nonlinear electrodynamics. This approach dates back to the 1930s with the work of Hoffmann and Infeld~\cite{Hoffmann:1937noa}, and received renewed attention in the early 2000s~\cite{Ayon-Beato:1998hmi,Ayon-Beato:2000mjt}. However, in these constructions the RBHs form a set of measure zero in the full solution space and the theories themselves possess a number of pathological properties~\cite{Bronnikov:2000vy,Bronnikov:2000yz,Fan:2016hvf,DeFelice:2024seu,Huang:2025uhv}. Importantly, constructions that obtain RBHs by coupling GR to matter fields actually contain all the singular solutions present in GR (as those solutions are recovered when the matter fields are turned off). Hence this is not a viable approach to RBH dynamics.

Very recently a new approach has emerged in which RBHs arise as exact solutions to theories that include infinite towers of higher--curvature corrections to GR~\cite{Bueno:2024dgm}. This is a compelling mechanism, as higher-curvature corrections are predicted by most quantum gravity frameworks~\cite{tHooft:1974toh,Goroff:1985th,Zwiebach:1985uq,Gross:1986mw,Grisaru:1986vi,Sakharov:1967nyk,Endlich:2017tqa,Eichhorn:2020mte,Borissova:2022clg}.\footnote{See \textit{e.g.} \cite{Bonanno:2000ep,Greenwood:2008ht,Saini:2014qpa,Kawai:2017txu, Harada:2025cwd} for other approaches to motivate the existence of RBHs.} The approach of~\cite{Bueno:2024dgm} employs \textit{quasi--topological gravities} (QT), which were originally constructed as gravitational and holographic toy models~\cite{Oliva:2010eb,Quasi,Dehghani:2011vu,Ahmed:2017jod,Cisterna:2017umf,Bueno:2019ycr, Bueno:2022res, Moreno:2023rfl,Moreno:2023arp}. While QT theories are engineered to have certain properties, such as second-order equations of motion in spherical symmetry, the resulting actions are not fine tuned in the space of theories. QT theories form a basis for the gravitational effective action in vacuum. Hence, the most general perturbative corrections to GR can be mapped to this class of theories~\cite{Bueno:2019ltp, Bueno:2019ycr, Bueno:2024dgm,Aguayo:2025xfi}. 

Remarkably, when an infinite tower of QT corrections is included in the action, singularity resolution occurs \textit{generically}, without any fine tuning of the coupling constants. Moreover, QT theories satisfy a Birkhoff theorem, ensuring the most general spherically symmetric solution is also static and characterized uniquely by the mass~\cite{Oliva:2010eb, Cisterna:2017umf, Bueno:2024eig, Bueno:2024zsx}. Hence, the RBH solutions are the unique solutions of the corresponding actions.  Altogether, this provides what is perhaps the first consistent approach in which dynamical questions about RBHs can be addressed. And indeed, working in this formalism it has been shown that the collapse of thin shells and stars of dust generically produces RBHs~\cite{Bueno:2024eig, Bueno:2024zsx,Bueno:2025gjg}. As such, it has attracted considerable recent attention~\cite{Konoplya:2024hfg, DiFilippo:2024mwm, Konoplya:2024kih, Ma:2024olw, Cisterna:2024ksz, Ditta:2024iky, Frolov:2024hhe,Wang:2024zlq, Fernandes:2025fnz, Fernandes:2025eoc,Hennigar:2025ftm,Aguayo:2025xfi,Cisterna:2025vxk,Boyanov:2025pes,Konoplya:2025uta, Eichhorn:2025pgy, Ling:2025ncw}.

In these models the study of RBH dynamics is facilitated by spherical dimensional reduction to two dimensions~\cite{Bueno:2024eig, Bueno:2024zsx,Bueno:2025gjg}. This reduction connects with earlier work on RBHs in two dimensions, where—unlike in higher dimensions—nontrivial dynamical analyses are tractable. Specific two-dimensional scalar–tensor theories admitting RBH solutions have been identified~\cite{Ziprick:2010vb, Kunstatter:2015vxa, Easson:2017pfe}, and several studies have used these models to investigate dynamical behaviour~\cite{Ziprick:2010vb, Biasi:2022ktq, Barenboim:2025ckx, Boyanov:2025pes}. All such investigations rely on two-dimensional Horndeski actions, which have been fully classified~\cite{Kunstatter:2015vxa, Carballo-Rubio:2025ntd}. In fact, \cite{Carballo-Rubio:2025ntd} provides a recipe by which one can begin with with a two-dimensional Horndeski action and then reverse engineer what the putative higher-dimensional equations of motion must have been. However, in any study that begins from the two-dimensional perspective it remains unclear whether the two-dimensional actions admitting RBH solutions can be obtained by dimensional reduction from fully covariant higher-dimensional theories.  The works~\cite{Bueno:2024dgm, Bueno:2024eig, Bueno:2024zsx,Bueno:2025gjg} overcome this in $D \ge 5$.

An important limitation of~\cite{Bueno:2024dgm} is that it applies only in $D \ge 5$ and not in the four-dimensional case relevant to our Universe. This is because QT theories exist only in five or more dimensions, and no four--dimensional theories of gravity with comparable computational control have been identified. On physical grounds it would be surprising to find that resolving singularities is \textit{easier} in higher dimensions than in four dimensions, since the strength of singularities typically increases with dimension. Moreover, it would be unexpected for a mechanism that operates for all $D \ge 5$ to fail specifically in $D = 4$. One may therefore expect that this limitation is a purely technical one, and with sufficient computational might it may still be possible to show that infinite towers of higher--curvature corrections can be resummed to obtain RBHs. It is the purpose of this work to explore this question in detail.

Let us review what is known about this problem in four dimensions.\footnote{Note, we exclude here any discussion of results which incorporate an \textit{ad hoc}/postulated matter. Such constructions do not provide a dynamical framework, since one has no idea how to generalize the \textit{ad hoc} stress tensor from one background to another.} We focus here on black hole singularities---in the homogeneous and isotropic cosmological setting there are four-dimensional models in which infinite towers of higher--curvature corrections can be resummed, and the resulting cosmologies have been shown to be nonsingular~\cite{Arciniega:2018tnn}. For black holes, arguably the most robust results come from nonminimal higher-derivative extensions of Einstein--Maxwell theory. In these constructions, an infinite tower of higher--derivative terms can produce actions which admit RBHs~\cite{Cano:2020qhy, Cano:2020ezi}. This construction has the desirable feature of allowing for the mass and charge to vary independently, avoiding the fine tuning constraints required in models based on nonlinear electrodynamics. The main limitation of this model is that matter is \textit{required} for singularity resolution, and the singular Schwarzschild metric is recovered for vanishing charge. 

RBHs and nonsingular cosmologies have recently been obtained from resummations of scalar--tensor extensions of GR ~\cite{Fernandes:2025eoc, Fernandes:2025fnz, Cisterna:2025vxk} based on four-dimensional regularizations of the Lovelock invariants~\cite{Lu:2020iav, Hennigar:2020lsl, Fernandes:2020nbq}. However, while it has been possible to establish the existence of regular asymptotically anti de Sitter black branes, the existence of regular asymptotically flat black holes or those with spherical horizon topologies remains inconclusive. Moreover, owing to the absence of a canonical kinetic term for the scalar, such models can suffer from strong coupling problems~\cite{Tsujikawa:2025eac}. Constructions generalizing this mechanism to vector--tensor theories have more recently appeared~\cite{Eichhorn:2025pgy} and offer the possibility of overcoming some of the classical inner horizon instabilities of RBHs \cite{Poisson:1990eh,Ori:1991zz,Frolov:2017rjz,DiFilippo:2022qkl,Barcelo:2022gii,Bonanno:2022jjp,Bonanno:2023qhp,Carballo-Rubio:2022pzu,Carballo-Rubio:2024dca,Arrechea:2024ajt,Bonanno:2025bgc}. However, the RBHs are not the general solutions of the corresponding action and require constraints between the integration constants. 

In each of the examples discussed above, additional dynamical fields besides the metric have been introduced. If one demands simultaneously four-dimensional RBHs from pure gravity and computational control, the only possibility is to consider actions non-polynomial in the curvature. While there is a vast literature on nonpolynomial theories,\footnote{For example, it is very common to consider $f(R)$ theories of gravity where the function $f(x)$ is not a polynomial.} to the best of our knowledge they were first applied to the problem of singularity resolution in~\cite{Mukhanov:1991zn}. In this, and several follow up studies~\cite{Mukhanov:1991zn, Brandenberger:1993ef, Brandenberger:1995hd, Yoshida:2017swb} it was shown that certain non-polynomial theories can yield nonsingular cosmologies and spherically symmetric black holes. More recently, these ideas have been independently developed for RBHs~\cite{Chinaglia:2017wim, Colleaux:2017ibe, Colleaux:2019ckh}, establishing classes of non-polynomial theories that satisfy a Birkhoff theorem in spherical symmetry and admit RBHs as exact solutions. 

In this work, we construct four-dimensional non-polynomial gravity theories with several desirable properties. Their field equations are second order on spherically symmetric and cosmological backgrounds and reduce to the naive four-dimensional limit of higher-dimensional quasi-topological gravities. We prove a Birkhoff theorem for these theories and show that, upon resumming an infinite tower of higher-curvature corrections, both black-hole and cosmological singularities are resolved. The main novelty of our work is a dynamical analysis via Oppenheimer–Snyder and thin-shell collapse, demonstrating that the RBH solutions of the theories are indeed formed by the collapse of matter. Altogether, this provides an explicit realization that the mechanism first identified in~\cite{Bueno:2024dgm} can apply equally well in four-dimensional spacetime. 

Our paper is structured as follows. First, we present the class of non-polynomial higher-curvature gravities we will be dealing with. We will show that their spherical sector reduces to two-dimensional Horndeski theories, thus featuring second-order equations on spherical symmetry. Also, we explain that their linearized spectrum on top of maximally symmetric background corresponds to a massless and traceless graviton, just like in Einstein gravity. Secondly, we demonstrate the mechanism by which these theories naturally produce both regular black holes and regular cosmologies by just imposing some mild conditions on the couplings determining the theory. Finally, we show how regular black holes form through dynamical collapse in two different popular scenarios in the literature: Oppenheimer-Snyder collapse and thin-shell collapse.

\section{NPQT gravities} 

In this section, we present non-polynomial higher-curvature theories that reduce to two-dimensional Horndeski theories on spherically symmetric backgrounds. For this reason, we will call these theories \emph{non-polynomial QTs} (NPQTs). Instances of non-polynomial QTs up to the fifth curvature order\footnote{If a non-polynomial higher-curvature theory is constructed from a linear combination of terms each of which has  dimension length$^{-2n}$, we will say the theory has curvature order $n$.} are given by:

\begin{widetext}

\begin{align}
\label{eq:z1}
\mathcal{Z}_{(1)}&=R\,, \\
\label{eq:z2}
\mathcal{Z}_{(2)}&=\frac{R^2}{12}+\frac{W_{abcd} W^{abcd}}{2}-Z_{ab}Z^{ab}\,, \\ \notag
\mathcal{Z}_{(3)}&=\frac{R^3}{144}+\frac{1}{8} R W_{ab}^{cd} W^{ab}_{cd}-\frac{1}{4} R Z_{a}^bZ^{a}_b+\frac{5}{2} W_{ab}^{cd}W_{cd}^{ef}W_{ef}^{ab}+3 W_{ab}^{cd}Z^{a}_c Z^{b}_d +Z_{a}^bZ_{b}^{c}Z_{c}^a  \\  & +\frac{9}{2}\frac{W_{ab}^{cd}W_{cd}^{ef}W_{ef}^{ab} Z_{g}^hZ_{h}^{i}Z_{i}^g W_{jk}^{lm} W_{lm}^{jk}}{W_{ab}^{cd} Z_c^a Z_d^b W_{ef}^{gh} W^{ef}_{gh}-2 W_{ab}^{cd}W_{cd}^{ef}W_{ef}^{ab} Z_g^h Z_h^g}  \,,  \label{eq:z3} \\ \notag
\mathcal{Z}_{(4)}&=\frac{R^4}{1728}+\frac{1}{48} R^2 W_{ab}^{cd} W^{ab}_{cd}-\frac{1}{24} R^2 Z_{a}^bZ^{a}_b+\frac{5}{6} R W_{ab}^{cd}W_{cd}^{ef}W_{ef}^{ab}+ R W_{ab}^{cd}Z^{a}_c Z^{b}_d+\frac{1}{3} R Z_{a}^bZ_{b}^{c}Z_{c}^a \\ \notag &    -\frac{9}{4} W_{ab}^{cd}W_{cd}^{ab} Z_e^f Z_f^e+21 W_{ab}^{cd} Z^a_c W^{eb}_{fd} Z_e^f+ \frac{1}{2}\left( W_{ab}^{cd} W_{cd}^{ab} \right)^2-3 Z_a^b Z_b^c Z_c^d Z_d^a+\frac{3}{2} \left( Z_a^b Z_b^a \right)^2\\&  \label{eq:z4} + \frac{3W_{ab}^{cd} W_{cd}^{ef}W_{ef}^{ab} Z_g^h Z_h^i Z_i^g}{W_{ab}^{cd} Z_c^a Z_d^b W_{ef}^{gh} W^{ef}_{gh}-2 W_{ab}^{cd}W_{cd}^{ef}W_{ef}^{ab} Z_g^h Z_h^g} \left ( 4 W_{hi}^{jk} Z_j^h Z_k^i+\frac{1}{2} R W_{hi}^{jk}W_{jk}^{hi} +6 W_{jk}^{hi}W_{hi}^{lm}W_{lm}^{jk}  \right)\,,\\
%+12  \frac{W_{ab}^{cd} Z_c^a Z_b^d W_{ef}^{gh} W_{gh}^{ij}W_{ij}^{ef} Z_k^l Z_l^m Z_m^k}{ W_{ab}^{cd} Z_c^a Z_d^b W_{ef}^{gh} W^{ef}_{gh}-2 W_{ab}^{cd}W^{cd}_{ef}W^{ef}_{ab} Z_g^h Z_h^g }+\frac{3}{2} \frac{R W_{ab}^{cd}W^{cd}_{ef}W^{ef}_{ab} Z_{g}^hZ_{h}^{i}Z_{i}^g W_{jk}^{lm} W_{lm}^{jk}}{W_{ab}^{cd} Z_c^a Z_d^b W_{ef}^{gh} W^{ef}_{gh}-2 W_{ab}^{cd}W^{cd}_{ef}W^{ef}_{ab} Z_g^h Z_h^g} 
%+\frac{3}{2} \frac{ \left (W_{ab}^{cd} W_{cd}^{ab} \right)^3 Z_e^f Z_f^g Z_g^e }{ W_{ab}^{cd} Z_c^a Z_d^b W_{ef}^{gh} W^{ef}_{gh}-2 W_{ab}^{cd}W^{cd}_{ef}W^{ef}_{ab} Z_g^h Z_h^g }\,, \label{eq:z4} \\
%+4 \frac{W_{ab}^{cd}W_{cd}^{ef}W_{ef}^{ab} Z_g^h Z_h^i Z_i^g}{W_{ab}^{cd} W_{cd}^{ab}}+\frac{2}{3} \frac{ \left (W_{ab}^{cd} W_{cd}^{ab} \right)^5 \left(Z_e^f Z_f^g Z_g^e \right)^3}{\left (W_{ab}^{cd} Z_c^a Z_b^d W_{ef}^{gh} W^{ef}_{gh}-2 W_{ab}^{cd}W^{cd}_{ef}W^{ef}_{ab} Z_g^h Z_h^g \right)^3 } 
\notag
\mathcal{Z}_{(5)}&=\frac{R^5}{20736}+\frac{5}{1728} R^3 W_{ab}^{cd} W^{ab}_{cd}-\frac{5}{864} R^3 Z_{a}^bZ^{a}_b+\frac{25}{144} R^2 W_{ab}^{cd}W_{cd}^{ef}W_{ef}^{ab}+\frac{5}{24} R^2 W_{ab}^{cd}Z^{a}_c Z^{b}_d +\frac{5}{72} R^2 Z_{a}^bZ_{b}^{c}Z_{c}^a   \\ \notag &+ \frac{5}{24} R \left (W_{ab}^{cd} W^{ab}_{cd} \right)^2-\frac{5}{4} R Z_a^b Z_b^c Z_c^d Z_d^a+\frac{5}{8} R \left( Z_a^b Z_b^a \right)^2-\frac{15}{16} R W_{ab}^{cd}W_{cd}^{ab} Z_e^f Z_f^e+\frac{35}{4} R W_{ab}^{cd} Z^a_c W^{eb}_{fd} Z_e^f\\ \notag &+ \frac{25}{4} W_{ab}^{cd} Z_c^a Z_d^b W_{ef}^{gh}W_{gh}^{ef}+\frac{65}{12} Z_a^b Z_b^a W_{cd}^{ef}W_{ef}^{gh}W_{gh}^{cd}+\frac{5}{6} Z_a^b Z_b^c Z_c^a W_{de}^{fg} W_{fg}^{de}+\frac{11}{12} W_{ab}^{cd} W_{cd}^{ab} W_{ef}^{gh}W_{gh}^{ij}W_{ij}^{ef} \\ \notag &+\frac{5 R W_{ab}^{cd}W_{cd}^{ef}W_{ef}^{ab} Z_{g}^hZ_{h}^{i}Z_{i}^g  }{W_{ab}^{cd} Z_c^a Z_d^b W_{ef}^{gh} W^{ef}_{gh}-2 W_{ab}^{cd}W_{cd}^{ef}W_{ef}^{ab} Z_g^h Z_h^g}  \left ( \frac{1}{16} R W_{jk}^{lm} W_{lm}^{jk} +\frac{3}{2}  W_{jk}^{lm} W_{lm}^{np} W_{np}^{jk} +  W_{jk}^{lm}Z_l^j Z_m^k\right)\\ \notag &+\frac{W_{ab}^{cd}W_{cd}^{ef}W_{ef}^{ab} Z_{g}^hZ_{h}^{i}Z_{i}^g }{W_{ab}^{cd} Z_c^a Z_d^b W_{ef}^{gh} W^{ef}_{gh}-2 W_{ab}^{cd}W_{cd}^{ef}W_{ef}^{ab} Z_g^h Z_h^g} \left (\frac{19}{6} \left ( Z_j^k Z_k^j \right)^2 +\frac{15}{4} \left (W_{jk}^{lm}W_{lm}^{jk} \right)^2-6 Z_j^k Z_k^l Z_l^m Z_m^j \right) \\ \nonumber &+\frac{W_{ab}^{cd}W_{cd}^{ef}W_{ef}^{ab}  }{W_{ab}^{cd} Z_c^a Z_d^b W_{ef}^{gh} W^{ef}_{gh}-2 W_{ab}^{cd}W_{cd}^{ef}W_{ef}^{ab} Z_g^h Z_h^g}  \left (\frac{55}{12}Z_{g}^hZ_{h}^{i}Z_{i}^g Z_h^i Z_i^h W_{jk}^{lm}W_{lm}^{jk}-5 W_{gh}^{ij} Z_i^g Z_j^h \left (Z_{k}^l Z_l^k \right)^2  \right)\\ \notag & +\frac{W_{ab}^{cd}W_{cd}^{ab} W_{ef}^{gh}Z_g^e Z_h^f }{W_{ab}^{cd} Z_c^a Z_d^b W_{ef}^{gh} W^{ef}_{gh}-2 W_{ab}^{cd}W_{cd}^{ef}W_{ef}^{ab} Z_g^h Z_h^g}  \left (\frac{25}{7}W_{ij}^{kl} Z_k^i Z_l^j Z_{m}^n Z_n^m+ \frac{1}{6}Z_{m}^n Z_n^p Z_p^m\left ( \frac{55}{2}W_{ij}^{kl}W_{kl}^{ij}+ Z_i^j Z_j^i\right)   \right)\\& +\frac{5\left (W_{ab}^{cd}W_{cd}^{ab} \right)^2}{W_{ab}^{cd} Z_c^a Z_d^b W_{ef}^{gh} W^{ef}_{gh}-2 W_{ab}^{cd}W_{cd}^{ef}W_{ef}^{ab} Z_g^h Z_h^g}  \left (\frac{1}{2} \left (Z_{e}^f Z_f^g Z_g^e \right)^2-\frac{2}{7} Z_{e}^f Z_f^e Z_g^h Z_h^i Z_i^j Z_j^g \right)  \,, \label{eq:z5}
\end{align}

\end{widetext}
where $W_{ab}^{cd}$ stands for the Weyl curvature tensor and $Z_{ab}=R_{ab}-\frac{1}{4}g_{ab}R$ is the traceless Ricci tensor. While \eqref{eq:z1} and \eqref{eq:z2} are free of non-polynomial terms --- in fact, they correspond to the Einstein-Hilbert term and the topological Gauss-Bonnet density, respectively ---, the remaining theories feature rational terms in which the denominator is always given by the same density: $W_{ab}^{cd} Z_c^a Z_b^d W_{ef}^{gh} W^{ef}_{gh}-2 W_{ab}^{cd}W_{cd}^{ef}W_{ef}^{ab} Z_g^h Z_h^g$. 

Despite the presence of these rational terms, we will prove in the following that these theories are indeed well defined on spherically symmetric backgrounds\footnote{The study of the non-polynomial terms appearing in these theories, including their potential singular (or nor) behavior for other type of backgrounds will be carried out elsewhere.} --- including Minkowski, de Sitter and anti-de Sitter.%, reducing to two-dimensional Horndeski theories.

\subsection{Non-polynomial QTs as two-dimensional Horndeski theories}

The key property ensuring that the theories \eqref{eq:z1} to \eqref{eq:z5} are well defined on spherical backgrounds is that their spherical sector is equivalent to two-dimensional Horndeski theories, just like the usual polynomial QTs \cite{Bueno:2024eig, Bueno:2024zsx}. To prove this, consider a general spherical configuration of the form:
\begin{equation}\label{sphericalmetric}
\mathrm{d}s^2=\gamma_{\mu\nu}\mathrm{d} x^{\mu}\mathrm{d}x^{\nu}+\varphi(x)^2 \mathrm{d}\Omega^2_{2}\, ,
\end{equation}
where $\gamma_{\mu\nu}\mathrm{d} x^{\mu}\mathrm{d}x^{\nu}$ stands for a two-dimensional Lorentzian metric (whose indices are denoted with Greek letters and run from 0 to 1) and $\varphi(x)$ is a scalar field that depends on the coordinates of the two-dimensional Lorentzian space. Observe that the four-dimensional tensors $W_{abcd}$,  $Z_{ab}$ and the Ricci scalar $R$  on top of \eqref{sphericalmetric} read:
\begin{align}
\label{eq:weyldec}
W_{ab}{}^{cd}&=\Omega \left[\gamma_{[a}{}^{c} \gamma_{b]}{}^{d}+\sigma_{[a}{}^{c} \sigma_{b]}{}^{d}  - \gamma_{[a}{}^{[c} \sigma_{b]}{}^{d]} \right] \,, \\
\label{eq:riccidec}
Z_{ab}&=\delta_{a}^\mu \delta_{b}^\nu \mathcal{S}_{\mu \nu}+\Theta \, \sigma_{ab}\,,\\
\label{eq:scaldec}
R&=R^{2{\rm d}}- \frac{4\Box \varphi}{\varphi}+2 \psi  \,,
\end{align}
where we introduced the orthogonal projectors\footnote{Note that $\gamma_a^b \gamma_b^c=\gamma_a^c$, $\sigma_a^b \sigma_b^c=\sigma_a^c$, $\gamma_a^b \sigma_b^c=0$, $\gamma_a^a =\sigma_a^a=2$.} $\gamma_{ab}=\delta_{a}^\mu \delta_{b}^\nu \gamma_{\mu \nu}$ and $\sigma_{ab}=g_{ab}-\gamma_{ab}$ , $R^{2{\rm d}}$ is the Ricci scalar of $\gamma_{\mu \nu}$, $\Box$ is the Laplacian associated with $\gamma_{\mu \nu}$ and:
\begin{align}
\label{eq:psiXdef}
\psi &=\frac{1-X}{\varphi^2}\, ,\quad X=\nabla_{\mu}\varphi\nabla^{\mu}\varphi\, , \\
\Omega&=\frac{2\psi}{3}+\frac{ R^{2{\rm d}}}{3}+\frac{2  \Box \varphi}{3 \varphi}\,, \quad 
\Theta=\frac{\psi}{2}-\frac{R^{2{\rm d}}}{4}\,,\\
\mathcal{S}_{\mu \nu}&= \left (\frac{R^{2{\rm d}}}{4}+\frac{\Box \varphi}{\varphi}- \frac{\psi}{2} \right) \gamma_{\mu \nu}  -\frac{2\nabla_\mu \nabla_\nu \varphi}{\varphi}  \,,
\end{align}
where $\nabla_{\mu}$ is the covariant derivative associated with $\gamma_{\mu\nu}$. Let us denote by $\mathcal{Z}_{(n)}^{2{\rm d}}$ with $n=1, \dots, 5$ the evaluation of the five theories \eqref{eq:z1} to \eqref{eq:z5} on top of the spherical ansatz \eqref{sphericalmetric}. Taking into account the previous results, one obtains:
\begin{align}
\label{eq:z1horn}
  &\hspace{-0.2cm} \mathcal{Z}_{(1)}^{2{\rm d}}=2 \psi-\frac{4 \Box \varphi }{\varphi}+R^{2{\rm d}}\,,  \\
   &\hspace{-0.2cm}\mathcal{Z}_{(2)}^{2{\rm d}}=2\psi R^{2{\rm d}}-\frac{8\mathfrak{t}_\varphi}{\varphi^2} \,, \\
   &\hspace{-0.2cm} \mathcal{Z}_{(3)}^{2{\rm d}}=6 \psi^3+\frac{12 \psi^2 \Box \varphi}{\varphi}+3 \psi^2 R^{2{\rm d}}-\frac{24 \psi  \mathfrak{t}_\varphi}{\varphi^2}\,,\\  &\hspace{-0.2cm}\mathcal{Z}_{(4)}^{2{\rm d}}=20 \psi^4+\frac{32 \psi^3 \Box \varphi}{\varphi}+4 \psi^3 R^{2{\rm d}}-\frac{48 \psi^2  \mathfrak{t}_\varphi}{\varphi^2}\,, \\
\label{eq:z5horn}
    &\hspace{-0.2cm}\mathcal{Z}_{(5)}^{2{\rm d}}=42 \psi^5+\frac{60 \psi^4 \Box \varphi}{\varphi}+5\psi^4 R^{2{\rm d}}-\frac{80 \psi^3 \mathfrak{t}_\varphi }{\varphi^2}\,,
\end{align}
where we have defined $\mathfrak{t}_\varphi=\nabla_a \partial^{[b} \varphi \nabla^{a]} \partial_b \varphi$. Observe that all rational terms with potentially diverging terms produce well-defined polynomial two-dimensional scalar-tensor theories. As a matter of fact, the theories \eqref{eq:z1horn} to \eqref{eq:z5horn} are instances of Horndeski theories \cite{Horndeski:1974wa}, with second-order equations of motion. As a result, the non-polynomial theories \eqref{eq:z1} to \eqref{eq:z5} possess second-order equations on top of spherical backgrounds, thus representing non-polynomial extensions of quasi-topological gravities to four dimensions. 

Now, let us use the following recursion formula to construct an infinite family of two-dimensional theories from the five theories \eqref{eq:z1horn} to \eqref{eq:z5horn}:
\begin{align}
\notag
    \mathcal{Z}_{(n+5)}^{2{\rm d}}&=\frac{(n+3) \mathcal{Z}_{(1)}^{2{\rm d}}\mathcal{Z}_{(n+4)}^{2{\rm d}}}{4(n+1)}-\frac{(n+4) \mathcal{Z}_{(2)}^{2{\rm d}}\mathcal{Z}_{(n+3)}^{2{\rm d}}}{4n}\\&+\frac{(n+3)(n+4) \mathcal{Z}_{(3)}^{2{\rm d}}\mathcal{Z}_{(n+2)}^{2{\rm d}}}{12n(n+1)}\,, \quad n\geq 1\,.
\end{align}
This recursive formula matches the four-dimensional extension of the higher-dimensional recursive formula satisfied by all polynomial quasi-topological gravities~\cite{Bueno:2019ycr,Bueno:2024zsx, Bueno:2024eig}. By direct induction, one can prove that the recursion formula is solved by:
\begin{align}
\notag
    \hspace{-0.2cm} \mathcal{Z}_{(n)}^{2{\rm d}}&=(2n-4)(2n-3)\psi^n+\frac{2n(2n-4)\psi^{n-1}\Box \varphi}{\varphi}\\& \hspace{-0.3cm} +n\psi^{n-1} R^{2{\rm d}}-\frac{4n(n-1) \psi^{n-2} \mathfrak{t}_\varphi}{\varphi^2}\,, \quad n \geq 1\,.
     \label{eq:znhorn}
\end{align}
The theories \eqref{eq:znhorn} belong to the Horndeski class for any $n\geq 1$. As a result, if we now go back to our four-dimensional theories \eqref{eq:z1} to \eqref{eq:z5} and consider the following higher-curvature theories  $\mathcal{Z}_{(n)}$ of curvature order $n>5$:
\begin{align}
\notag
    \mathcal{Z}_{(n+5)}&=\frac{(n+3) \mathcal{Z}_{(1)}\mathcal{Z}_{(n+4)}}{4(n+1)}-\frac{(n+4) \mathcal{Z}_{(2)}\mathcal{Z}_{(n+3)}}{4n}\\&+\frac{(n+3)(n+4) \mathcal{Z}_{(3)}\mathcal{Z}_{(n+2)}}{12n(n+1)}\,, \quad n \geq 1\,,
    \label{eq:recformula}
\end{align}
it follows that the corresponding  $\mathcal{Z}_{(n)}$ for $n >5$ will produce two-dimensional Horndeski theories when evaluated on the spherically symmetric background~\eqref{sphericalmetric}. Consequently, \eqref{eq:recformula} provides a way to construct non-polynomial QTs at any curvature order from the five theories \eqref{eq:z1} to \eqref{eq:z5}. Crucially, the rational terms present in the four-dimensional theories possess a well-defined spherically symmetric limit and produce polynomial terms, which combined with the rest of higher-curvature terms give rise to two-dimensional Horndeski theories. 

By virtue of these results, we may consider four-dimensional theories constructed from infinite towers of non-polynomial QTs:
\begin{equation}
\label{eq:QTaction}
    S=\frac{1}{16 \pi G_N} \int \mathrm{d}^4 x \sqrt{\vert g \vert} \left[ R+\sum_{n=2}^\infty \alpha_n \mathcal{Z}_{(n)} \right ]\,,
\end{equation}
where the couplings $\alpha_n$ possess dimensions of length$^{2n-2}$ and are completely free. The inclusion of an infinite number of terms will be crucial for singularity resolution, as will become manifest later. However, one could truncate the series at a finite order (if desired) and study the subsequent theory with a finite number of curvature terms --- featuring singular solutions, though.

Performing a dimensional reduction of \eqref{eq:QTaction} on \eqref{sphericalmetric}, one ends up with the following two-dimensional Horndeski theory:
\begin{equation}\label{eq:2daction}
S_{\rm 2d}=\frac{1}{2 G_{\rm N}}\int \mathrm{d}^{2}x\sqrt{|\gamma|} \mathcal{L}_{\rm 2d}(\gamma_{\mu\nu},\varphi)\, ,
\end{equation}
where
\begin{align}
\label{eq:horn}
&\mathcal{L}_{\rm 2d}=G_{2}(\varphi, X)-\Box\varphi G_{3}(\varphi, X)+G_{4}(\varphi, X)R^{\rm 2d}
\nonumber \\
&-2G_{4,X}(\varphi, X)\left[(\Box\varphi)^2-\nabla_{\mu}\nabla_{\nu}\varphi\nabla^{\mu}\nabla^{\nu}\varphi\right]\, ,
\end{align}
having defined the functions $G_i(\varphi,X)$ as follows:
\begin{align}
G_{2}(\varphi, X)&=\varphi^{2}\left[3 h(\psi)-2\psi  h'(\psi)\right]\, ,\\\label{G3form}
%-2\psi^{(D+1)/2}\frac{\mathrm{d}}{\mathrm{d}\psi}\left[\psi^{(1-D)/2}h(\psi)\right]\, ,\\
G_{3}(\varphi, X)&=2\varphi h'(\psi)\, ,\\
\label{eq:G4}
G_{4}(\varphi, X)&=-\frac{1}{2}\varphi^{2}\psi \int \mathrm{d}\psi \psi^{-2}h'(\psi)\, ,
\end{align}
Observe the notation  $G_{i,X} \equiv \partial_X G_i$ (similarly, $G_{i,\varphi} \equiv \partial_\varphi G_i$, to be used later). All the information about the 4-dimensional theory \eqref{eq:QTaction} is now encoded in the \emph{characteristic function} $h(\psi)$, 
 \be \label{eom_psi}
h(\psi)\equiv \psi + \sum_{n=2}^{\infty} (2-n) \alpha_n   \psi^n\, .
\ee

%\comment{HEREEEE}\\
All the expressions above coincide with those reported in \cite{Bueno:2024eig, Bueno:2024zsx,Bueno:2025gjg} for standard (polynomial) quasi-topological gravities in arbitrary $D\geq 5$, could one allow for $D=4$ in that context --- which is not permitted. Hence, both the analysis of the vacuum solutions, including the existence of a Birkhoff theorem, as well as the problem of gravitational collapse for spherical matter may be now adapted to four dimensions, by virtue of the NPQTs presented in this document. This will be addressed in Sections \ref{sec:rbh} and~\ref{sec:col}.

\subsection{Linearized spectrum on maximally symmetric backgrounds}

Let us now show that non-polynomial QTs only propagate the usual massless graviton of Einstein gravity on top of maximally symmetric backgrounds. To this aim, let us split the space-time metric $g_{ab}$ into:
\begin{equation}
   g_{ab}=\bar{g}_{ab}+\varepsilon h_{ab}\,,
\end{equation}
where $\bar{g}_{ ab}$ is a maximally symmetric space-time (Minkowski, de Sitter or anti-de Sitter), $h_{ab}$ stands for the perturbation and $\varepsilon$ is to be regarded as a book-keeping parameter. Observe that:
\begin{align}
\nonumber
    W_{abcd}&=\varepsilon \, W_{abcd}^{(1)}+ \mathcal{O}(\varepsilon^2)\,,  \quad Z_{a b}=\varepsilon Z_{ab}^{(1)}+ \mathcal{O}(\varepsilon^2)\,,\\
    R&=\bar{R}+ \varepsilon R^{(1)} +\varepsilon^2 R^{(2)}+\mathcal{O}(\varepsilon^3)\,,
\end{align}
where $\bar{R}$ denotes the constant scalar curvature of $\bar{g}_{ab}$. As a result, if we start from a $n$-th order non-polynomial QT $\mathcal{Z}_{(n)}$, the quadratic theory from which the linearized equations for $h_{ab}$ may be found is only sensitive\footnote{For this procedure to be well defined, the perturbations $h_{ab}$ must be such that the rational terms in the Lagrangians do not diverge.} to the $R^n$, $R^{n-2} W_{ab}^{cd} W^{ab}_{cd}$ and $R^{n-2}Z_{a}^{b} Z_b^{a}$ pieces in $\mathcal{Z}_{(n)}$. Therefore, we may just focus on these terms. Remarkably, by use of \eqref{eq:recformula} one may show that:
\begin{align}
\notag
    \mathcal{Z}_{(n)}=\frac{R^{n}}{12^{n-1}} &\left[ 1+\frac{3n(n-1)}{R^2} \left( W_{ab}^{cd} W^{ab}_{cd}-2 Z_{a}^b Z_b^a\right) \right.\\ \label{eq:zcuad} &+ \left. \mathcal{O}\left ( R^{-3} \right) \right]\,,
\end{align}
Interestingly enough, the coefficients affecting the terms $R^n$, $R^{n-2} W_{ab}^{cd} W^{ab}_{cd}$ and $R^{n-2}Z_{a}^{b} Z_b^{a}$ are exactly the same coefficients that multiply the same terms in the four-dimensional generalized quasi-topological theories identified in \cite{Moreno:2023rfl} for any curvature order $n$. Since these theories possess the same linearized spectrum on top of maximally symmetric backgrounds as Einstein gravity, we conclude that the non-polynomial QT gravities presented in this document will only propagate a massless graviton when perturbed around maximally symmetric space-times.

% Indeed, there are hints that the principle of ``infinite tower $\Rightarrow$ singularity resolution'' is more general coming from other approaches~\cite{Fernandes:2025fnz, Fernandes:2025eoc}. 

% It is the purpose of 

% It is the purpose of this work to present a class of four-dimensional gra

% The resummed QT theories satisfy a scaling property that enforces a universal (mass independent) upper bound on the curvature, manifesting Markov's limiting curvature hypothesis~\cite{PismaZhETF.36.214, Frolov:2024hhe, Bueno:2024zsx}.

%\comment{Emphasize what happens with the denominator pieces when the dimensional reduction is performed}

\subsection{Equations of motion}

The complete set of gravitational equations of motion for our four-dimensional non-polynomial QTs take the form:
\begin{equation}
    \mathcal{E}_{ab}=8 \pi G_N T_{ab}\,, 
    \label{eq:eomgen}
\end{equation}
where we have defined
\begin{equation}
    \mathcal{E}_{ab}=\frac{16 \pi G_N}{\sqrt{\vert g \vert}} \frac{\delta S}{\delta g^{ab}}
\end{equation}
and $T_{ab}$ is the matter stress-energy tensor. We will be interested in evaluating the equations \eqref{eq:eomgen} on spherical backgrounds \eqref{sphericalmetric}. Resorting to the equivalent two-dimensional Horndeski picture provided by \eqref{eq:horn}, one easily finds that the unique non-zero components of the equations of motion on \eqref{sphericalmetric} are given by:
\begin{align}
\mathcal{E}_{\mu \nu}&=\frac{2G_3}{\varphi^2} g_{\mu[\nu} \nabla_{\beta]} \partial^\beta \varphi-\frac{G_2}{\varphi^2} g_{\mu \nu}\,, \\
  \mathcal{E}_{ij}&=\frac{g_{ij}}{2\varphi} \left[ G_{3,\varphi} \Box \varphi-G_{2,\varphi}-\frac{G_3}{2} R-2 G_{3,X} \mathfrak{t}_\varphi \right]\,,
\end{align}
where $\mu, \nu$ denote two-dimensional components and the indices $i,j$ correspond to angular directions. Observe that the angular components are pure gauge and may be obtained from the two-dimensional ones \cite{Bueno:2024zsx,Bueno:2025gjg}.  We will now further specify our spherical ansatz \eqref{sphericalmetric} to describe either black holes solutions or Friedmann-Lema\^itre-Robertson-Walker (FLRW) cosmologies.

\subsubsection{Black hole space-times}

Let us fix the gauge $\varphi=r$ and set

%Let us find the equations of motion on spherical symmetry of a general non-polynomial QT, possibly including an infinite tower of higher-curvature terms. 

%To this aim, we are going to fix the gauge $\varphi=r$ and set
%The equations of motion in the spherically symmetric sector can be obtained from the effective two-dimensional action \eqref{eq:horn}. 
%As a matter of fact, these equations take the same form as those shown in  \cite{ Bueno:2024zsx,Bueno:2025gjg}, the equation for $\varphi$ is pure gauge and one can set
\begin{equation}
\mathrm{d}s_\gamma^2=-N(t,r)^2 f(t,r) \mathrm{d}t^2+\frac{\mathrm{d}r^2}{f(t,r)} \, , \quad \varphi=r
\label{eq:ssans2}
\end{equation}
without loss of generality. Following the same formal steps as in \cite{ Bueno:2024zsx,Bueno:2025gjg} --- setting $D=4$ ---, one infers that the whole set of gravitational equations of motion boils down to:
\begin{align}
\label{eom1}
 \partial_r \left[r^{3}h(\psi) \right]&=\frac{8\pi G_{\rm N}}{N^2 f} r^{2}T_{tt} \,,\\
 \label{eom2}
\partial_t f  &=-\frac{8\pi G_{\rm N}}{h'(\psi)} rf T_{tr} \,  \,,\\
\label{eom3}
\partial_r N &=\frac{4\pi G_{\rm N} }{h'(\psi)} r N\left(T_{rr}+\frac{1}{N^2 f^2}T_{tt}\right) \,.
\end{align}
where $T_{ab}$ is the (four-dimensional) matter stress tensor. Now, $X=f(t,r)$, so that $r^2\psi=1-f(t,r)$.
%\be \label{eom}
%h(\psi)  = \frac{2 G_{\rm N} M }{r^3}  \, ,\quad 
%\psi\equiv \frac{1-f(r)}{r^2}\, ,
%\ee

\,

{\bf Equations in vacuum}. For $T_{ab}=0$,  Eqs.\,(\ref{eom1})-(\ref{eom3}) simply reduce to
 \begin{equation}\label{vacuumeq}
 \partial_t f=0\, , \quad \partial_r N=0\, , \quad  \frac{\partial}{\partial r} \left[r^{3}h(\psi) \right]=0\, .
 \end{equation}
 Hence, $N=N(t)$ can be removed by redefining the time coordinate and $f=f(r)$.  Thus, we find that the most general spherically symmetric solution of a general NPQT theory is static and of the form
 %$N(t)^2 {\mathrm d}t^2 \rightarrow  {\mathrm d}t^2$. We thus conclude that the most general spherically symmetric solution of \req{QTaction} is in fact static and fully determined by a single function $f(r)$,
 \begin{equation}
\mathrm{d}s^2=- f(r) \mathrm{d}t^2+\frac{\mathrm{d}r^2}{f(r)}+r^2 \mathrm{d}\Omega_2^2\,.
\label{eq:ssans}
\end{equation}
Integrating the last equation in \req{vacuumeq} yields an algebraic equation for $f(r)$, namely,
 \be \label{eomh}
h(\psi)  = \frac{2G_N M}{r^{3}}  \, ,%\quad \psi\equiv \frac{1-f(r)}{r^2}\, ,
\ee
where the integration constant $M$ is the ADM mass  \cite{Arnowitt:1960es,Arnowitt:1960zzc,Arnowitt:1961zz,Deser:2002jk} of the solution. %, $M$, through 
%\begin{equation}\label{newM}
%\mathsf{M} \equiv \frac{8\pi G M}{(D-2)\Omega_{(D-2)}}\, . 
%\end{equation}
Hence, the NPQT theories satisfy a Birkhoff theorem.
%This proves that our theories satisfy a Birkhoff theorem, extending previous results in the literature \cite{Oliva:2010eb, Oliva:2011xu, Cisterna:2017umf}. %Next, we study in detail the properties of the solutions. 

\subsubsection{Cosmological backgrounds}

Consider now FLRW space-times with spherical sections. This is captured by the general spherical metric \eqref{sphericalmetric} after setting
\begin{equation}
    \mathrm{d}s^2_\gamma=-\mathrm{d}\tau^2+\frac{a(\tau)^2 d\eta^2}{1-\eta^2}
\end{equation}
and $\varphi=a(\tau) \eta$, where $a(\tau)$ stands for the scale factor.  Assume the presence of a perfect fluid with density $\rho$, pressure $p$ and four-velocity $u^a=\delta_\tau^a$. The associated stress-energy tensor is given by
\begin{equation}
    T_{ab}=(\rho+p) u_a u_b+p g_{ab}\,.
\end{equation}
Then, the Friedmann equation for  $a(\tau)$ reads:
\begin{equation}
    h(\Phi)=\frac{8\pi G_N}{3} \rho\,, \quad \text{where }\,  \Phi=\frac{1+\dot{a}(\tau)^2}{a(\tau)^2}
    \label{eq:friedqts}
\end{equation}
and $\dot{a}$ is the time derivative of the scale factor. Similarly, the conservation equation for the fluid takes the form:
\begin{equation}
    \dot{\rho}+3(\rho+p) \frac{\dot{a}}{a}=0\,.
    \label{eq:conteq}
\end{equation}
If we consider a linear barotropic equation of state for the fluid, so that
\begin{equation}
    p=w \rho\,, 
    \label{eq:eos}
\end{equation}
the continuity equation can be exactly integrated:
\begin{equation}
    \rho=\frac{3{\sf a}}{8 \pi G_N a(\tau)^{3(w+1)}}\,,
\end{equation}
and substituting back into the Friedmann equation \eqref{eq:friedqts}, one finds
\begin{equation}
    h(\Phi)=\frac{{\sf a}}{ a(\tau)^{3(w+1)}}\,.
    \label{eq:cosmonpqt}
\end{equation}
As for spherically symmetric black holes, cosmological backgrounds are also fully determined by the characteristic function $h$ that fixes the particular tower of non-polynomial QTs under consideration. Note the formal equivalence of \eqref{eq:cosmonpqt} with \eqref{eomh} under the exchanges of $\psi \leftrightarrow \Phi$, $2 G_N M \leftrightarrow \mathsf{a}$ and $r \leftrightarrow a(\tau)^{w+1}$.

%\comment{tbe}

\section{Regular black holes and regular cosmologies}

\label{sec:rbh}

In this section, we will show that the inclusion of an infinite tower of non-polynomial QTs will generically lead to the resolution of singularities for both black hole solutions and FLRW space-times, as long as some relatively mild conditions on the couplings that specify the theory are met. We will elaborate first on regular black holes, and then explain that the same theories automatically allow for regular cosmologies.

\subsubsection{Regular black holes}

Just like in the case of higher-dimensional QT theories, the Schwarzschild black hole singularity is weakened by the presence of NPQT terms. By direct study of \eqref{eomh}, one observes that if one only includes non-polynomial QTs up to some finite order $n=n_{\rm max}$, then
\be 
f (r)= 1 - \left(\frac{2\mathsf{M} }{\alpha_{n_{\rm max} }} \right)^{\frac{1}{n_{\rm max}} } r^{2  - \frac{3}{n_{\rm max}} } + \cdots \, ,
\ee
near $r=0$. Remarkably, when an infinite tower of corrections is included, the singularity is completely resolved, as long as the function $h(\psi)$ has an inverse for $\psi>0$ and the series that defines it via Eq.~\req{eom_psi} has a finite radius of convergence. These conditions are satisfied by very general sets of the $\alpha_n$ couplings. In particular, the following\footnote{If necessary, such as for the choice of couplings  \eqref{coup}, $\lim_{n\rightarrow\infty}$ is to be replaced by $\limsup_{n\rightarrow\infty}$.} are sufficient conditions \cite{Bueno:2024eig}
\begin{equation}
\label{eq:condcoup}
    \alpha_{n}(2-n)\ge 0\,\, \forall\, n\, , \quad \text{and} \quad \lim_{n\rightarrow\infty} |\alpha_{n}|^{\frac{1}{n}}>0\, .
\end{equation}
 % Observe that $\alpha_n$ has dimensions of length$^{2(n-1)}$. Hence, if we make explicit the length scale, $\alpha_n \equiv \bar\alpha_n \ell^{2(n-1)}$,
%  the second condition simply means that the dimensionless coefficients tend to a constant value as $n\rightarrow \infty$.
Naturally, different choices of the $\alpha_n$ represent different NPQT theories which possess different black hole solutions. For the sake of concreteness, we will focus on a particular model which yields simple expressions and allows for analytic calculations. In odd dimensions, a particularly suitable choice yields a $D$-dimensional version of the Hayward black hole \cite{Hayward:2005gi}. On the other hand, the fact that in even dimensions the density of order $n=D/2$ is 
topological does not allow for such a solution. In our case, $\mathcal{Z}_{(2)}$ is nothing but the Gauss-Bonnet invariant. In $D=4$ this makes no contribution to the equations of motion, implying that the quadratic term is absent from $h(\psi)$. 

\begin{figure}[t!]
\centering 
\includegraphics[width=0.47\textwidth]{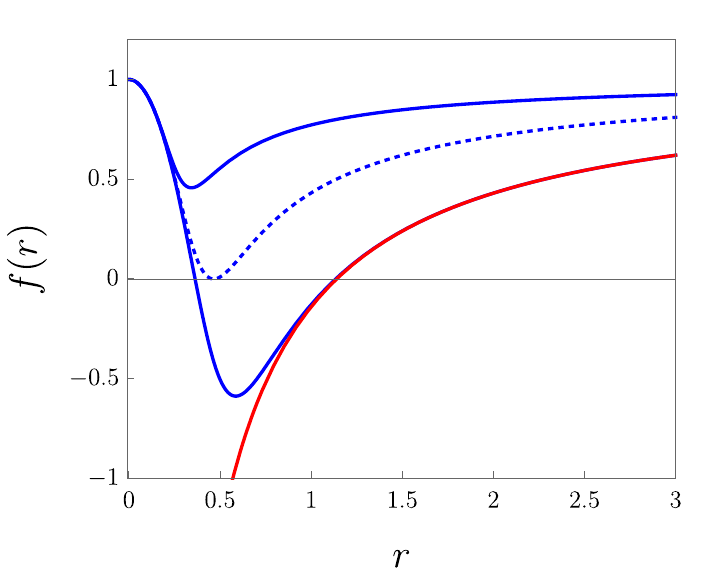}
\caption{We plot $f(r)$ for the unique spherically symmetric vacuum solution of a NPQT theory with infinitely many higher-derivative terms with couplings given by Eq.\,(\ref{coup}) and $\alpha=1/8$. The lower blue curve corresponds to a regular black hole with two horizons ($M=2M_{\rm cr}$). The red curve is the usual Schwarzschild black hole metric function with the same mass. The intermediate dotted blue curve represents an extremal black hole with a single horizon ($M=M_{\rm cr}$). The upper blue curve corresponds to a horizonless regular spacetime ($M=\frac{5}{8}M_{\rm cr}$).  }
\label{fig:modiH}
\end{figure}

An alternative choice of couplings  which yields a simple uniparametric family of solutions in $D=4$ corresponds to the characteristic polynomial
\begin{equation}
    h(\psi)=\frac{\psi}{\sqrt{1-\alpha^2\psi^2}} \, , \quad \alpha>0\,,
\end{equation}
which entails choosing
\begin{equation}\label{coup}
    (2-n)\alpha_n=\frac{[1-(-1)^n]\Gamma(\frac{n}{2})}{2\sqrt{\pi}\Gamma\left(\frac{n+1}{2} \right)} \alpha^{n-1}\, .
\end{equation}
For this, the metric function reads
\begin{equation}
    f(r)=1-\frac{2G_N M r^2}{\sqrt{r^6+(2G_N M\alpha)^2}}\, .
\end{equation}
The solution reduces to the Schwarzschild metric asymptotically, but has a regular de Sitter core instead of a curvature singularity, namely,
\begin{align}
     f(r)&\overset{(r\gg 2G_N M)}{\sim} 1-\frac{2G_N M}{r}+\frac{4\alpha^2(G_N M)^3}{r^7}+\dots\, ,  \\ f(r)&\overset{(r\ll 2G_N M)}{\sim} 1-\frac{r^2}{\alpha}+\frac{r^8}{8\alpha^3 (G_N M)^2}+\dots\, 
\end{align}
   Whenever 
   \begin{equation}
      M>M_{\rm cr}\equiv \frac{3^{3/4}}{2^{3/2}G_N} \sqrt{\alpha}\, ,
   \end{equation}
   the solution describes a black hole with two horizons. It becomes extremal for $M=M_{\rm cr}$, whereas for $M<M_{\rm cr}$ it describes a globally regular horizonless spacetime --- see Fig.\,\ref{fig:modiH}.

In all cases the curvature invariants remain finite everywhere and, remarkably, they possess a mass-independent maximum value, in agreement with Markov's limiting curvature hypothesis \cite{Markov}. In fact, this feature has been proven to hold in general for all $D\geq 5$ QT theories in \cite{Frolov:2024hhe}. The argument relies exclusively on the existence of the characteristic polynomial $h(\psi)$ and the structure of the $D$-dimensional version of Eq.\,(\ref{eomh}), so it extends straightforwardly to general NPQT black holes. In the particular case of a theory with couplings given by Eq.\,(\ref{coup}), the Kretschmann invariant is bounded above by 
\begin{equation}
    R_{abcd}R^{abcd}\leq \frac{24}{\alpha^2}\, , 
\end{equation}
which is explicitly independent of the mass of the solutions. For this particular model, the maximum is always reached at $r=0$, but this is not the case in general \cite{Bueno:2024zsx}.

\subsubsection{Regular cosmologies}

The sufficient conditions \eqref{eq:condcoup} to produce regular black hole configurations also guarantee the absence of singularities in FLRW space-times. As a matter of fact, it is quite direct to show that the dynamical evolution for the scale factor $a(\tau)$ is bounded and take place as a sequence of expansions and contractions, never reaching a vanishing scale factor.

To this aim, let us consider an infinite tower of non-polynomial QTs in which the couplings satisfy the conditions \eqref{eq:condcoup}. Assume the presence of a perfect fluid with a linear barotropic equation of state \eqref{eq:eos}, so that the evolution equation for $a(\tau)$ is determined by \eqref{eq:cosmonpqt}. Consider a sequence of couplings $\{ \alpha_n \}_{n=2}^\infty$ such that $\lim_{n\rightarrow\infty} |\alpha_{n}|^{\frac{1}{n}}=C>0$. Following the argumentation of \cite{Bueno:2025gjg}, assuming that $w >-1$ it can be shown that the regime of small scale factor is governed by
\begin{equation}
    \dot{a}(\tau)^2+1=\frac{a(\tau)^2}{C}\,,
\end{equation}
whose solution reads
\begin{equation}
    a(\tau)=\sqrt{C}\cosh \left (\frac{\tau-\tau_0}{\sqrt{C}} \right)\,,
\end{equation}
where we have set as a initial condition that $\dot{a}(\tau_0)=0$, for certain initial time $\tau_0$. We clearly see that this corresponds to a bouncing behavior\footnote{In the cosmologies provided by the theory with couplings \eqref{coup}, one should proceed more carefully, as $\lim_{n\rightarrow\infty} |\alpha_{n}|^{\frac{1}{n}}$ does not exist --- it is $\limsup_{n\rightarrow\infty} |\alpha_{n}|^{\frac{1}{n}}$ that is well defined. However, the qualitative aspects do not change and one also gets a bouncing evolution for the scale factor, see Figure \ref{fig:modiH0}.} for $a(\tau)$, in which the scale factor attains a non-vanishing minimum value of $a(\tau)=\sqrt{C}$.

In another vein, it can also be shown that the scale factor is also bounded from above. Since this is a pure IR effect, higher-curvature terms may be neglected and one may study this regime by just considering the Einstein-Hilbert term in \eqref{eq:QTaction}. In this limit, \eqref{eq:cosmonpqt} becomes:
\begin{equation}
    1+\dot{a}(\tau)^2=\mathsf{a}\, a(\tau)^{-1-3w}\,.
    \label{eq:maxa}
\end{equation}
Examining this equation around an extremum of $a(\tau)$ located $\tau=\tau_0$, so that  $\dot{a}(\tau_0)=0$, we may expand $a(\tau)=a_0+a_2 (\tau-\tau_0)^2+\dots$. Substituting on \eqref{eq:maxa}, one gets:
\begin{equation}
    a_0=\mathsf{a}^{\frac{1}{1+3w}}\,, \quad a_2=-\frac{(1+3w)}{4 a_0}\,,
\end{equation}
so that whenever\footnote{Note that $w < -1/3$ violates the strong energy condition and requires inflationary matter.} $w>-\frac{1}{3}$, the evolution of $a(\tau)$ indeed corresponds to a bounce around the maximum $a(\tau_0)=a_0$. For large values of the scale factor, it is the Einstein-Hilbert term alone which halts the growth of $a(\tau)$, while the regular dynamics for small values of the $a(\tau)$ is ensured by the infinite tower of higher-curvature corrections. This behavior is corroborated in Figure \ref{fig:modiH0}, which plots the radius of a pressureless dust star in an Oppenheimer-Snyder collapse as predicted by the non-polynomial QT theory with couplings \eqref{coup} --- note that this radius is proportional to the scale factor, cf.  \eqref{ratu}.

%\begin{figure}[t!]
%\centering 
%\includegraphics[width=0.47\textwidth]{Evol_Cosmo.pdf}
%\caption{Time evolution of the scale factor $a(\tau)$ in a dust model $w=0$. In red, we plot the scale factor in GR, which approaches the singularity near $\tau \sim 1.45$ --- using units of $\mathsf{a=1}$. In blue, we present the scale factor for the theory provided by the couplings \eqref{coup} with $\alpha=1/4$. Interestingly, we observe that the scale factor remains bounded and undergoes a periodic motion featured by infinite bounces.} %We plot $f(r)$ for the unique spherically symmetric vacuum solution of a NPQT theory with infinitely many higher-derivative terms with couplings given by Eq.\,(\ref{coup}) and $\alpha=1/8$. The lower blue curve corresponds to a regular black hole with two horizons ($M=2M_{\rm cr}$). The red curve is the usual Schwarzschild black hole metric function with the same mass. The intermediate dotted blue curve represents an extremal black hole with a single horizon ($M=M_{\rm cr}$). The upper blue curve corresponds to a horizonless regular spacetime ($M=\frac{5}{8}M_{\rm cr}$).  }
%\label{fig:cosmoMod}
%\end{figure}

\section{Gravitational Collapse}
\label{sec:col}
We turn now to the problem of gravitational collapse. We first review how the modified junction conditions can be obtained from the effective two-dimensional action, adapting the $D\geq 5$ results of \cite{Bueno:2024zsx,Bueno:2024eig}. Then, we analyze two simple collapse processes corresponding to pressureless matter: a dust star and a thin shell.

\subsection{Modified junction conditions}
The modified Israel junction conditions \cite{Israel:1966rt}  required for the study of spherical collapse of matter in $D$-dimensional QT models whose two-dimensional effective action is of the Horndeski type (\ref{eq:horn}) were obtained in \cite{Bueno:2024zsx}. These can be immediately adapted to the present case by simply setting $D=4$ in the corresponding formulas. Let us present here the relevant results.

Consider a $4$-dimensional manifold divided in two, $\mathcal{M}_+$ and $\mathcal{M}_-$, by some interface $\Sigma$, where some stress-energy tensor $S_{AB}$ may be localized (we use capital Latin indices to denote boundary indices). The first junction condition is identical to the one in general relativity, namely, the induced metrics on $\Sigma$ obtained from $\mathcal{M}_+$ and $\mathcal{M}_-$ must match,
\begin{equation}
h_{AB}^+=h_{AB}^-\,.
\end{equation}
Assuming this condition holds, we can simply denote $h_{AB}=h_{AB}^+=h_{AB}^-$.

Now, let $n_a$ be a unit normal vector to $\Sigma$ normalized so that $n_a n^a= \varepsilon=\pm 1$. Assuming spherically symmetry, we can write 
\begin{equation}
h_{AB} \mathrm{d}x^A \mathrm{d}x^B=h_{\tau \tau} \mathrm{d}\tau^2+ \varphi(\tau)^2 \mathrm{d}\Omega_{2}^2\,.
\end{equation} 
Relative to the two-dimensional metric $\gamma_{\mu\nu}$, $\Sigma$ is just a curve whose induced metric reads $\mathrm{d}s^2_h=h_{\tau \tau} \mathrm{d}\tau^2$. The proper time of the curve can be chosen to be $\tau$, so that one can always fix the gauge $h_{\tau \tau}=-\varepsilon$. We shall do this later in our examples. %Let us keep $h_{\tau \tau}$ general for now in order to analyze the variational problem of the two-dimensional theory.
%If we choose $\tau$ to be the proper time of the curve, one can always set $h_{\tau \tau}=-\varepsilon$. However, it will be convenient to keep $h_{\tau \tau}$ general for the moment, in order to examine properly its variational problem. 

The second junction condition does depend on the theory under consideration and in general differs from the general relativity one. In order to determine it, we need the boundary terms which give rise to a well-posed variational problem for the theory under consideration. Restricted to the spherically symmetric sector, the problem reduces to obtaining the boundary term for the Horndeski theory which follows from the dimensional reduction. The result can be found in \cite{Bueno:2024zsx}, which agrees with previous results for general Horndeski theories \cite{Padilla:2012ze}. The full two-dimensional action reads $S_{\rm 2d}^{\rm tot.}=S_{\rm 2d}+S_{\rm 2d}^{\rm bdry.}$, where
%This condition relies heavily on the analytical derivation of those boundary terms that need to be added to \eqref{QTaction} to produce a well-posed variational problem. While the direct finding of these terms from the $D$-dimensional action \eqref{QTaction} is quite problematic, it becomes a feasible task when restricting to spherical symmetry and using the equivalent two-dimensional theory. Indeed, the total two-dimensional action with the appropriate boundary terms that gives rise to a well-posed two-dimensional variational problem is \cite{Padilla:2012ze,Bueno:2024eig,Bueno:2024zsx}:
\begin{equation}
S_{\rm 2d}^{\rm bdry.}= \int_\Sigma  \frac{ \mathrm{d}\tau \sqrt{\vert h_{\tau \tau} \vert} }{2 G_N}\left[F_3+2G_4 K+4 \Box^h \varphi \, F_{4,Y} \right]\,,
\label{eq:2dtot}
\end{equation}
where $K\equiv\nabla_\mu n^\mu$, $\Box^h$ is the Laplacian operator on $\Sigma$ and
\begin{equation}
F_l\equiv \int_0^{n^\mu \partial_\mu \varphi} G_l(\varphi, Y+\varepsilon x^2) \, \mathrm{d}x\,, \quad Y\equiv h^{\tau \tau} \dot{\varphi}^2\,, 
\end{equation}
where $l=3,4$, $\dot{\varphi}\equiv \mathrm{d}\varphi/\mathrm{d}\tau$ and $h^{\tau \tau}=1/h_{\tau \tau}$. The boundary equations of motion read
\begin{equation}
    \Pi_{AB}\equiv \frac{16\pi G_N}{\sqrt{|h|}}\frac{\delta S_{2d}^{\rm tot.}}{\delta h_{AB}}\, .
\end{equation}
In terms of these, the second junction condition for a NPQT in the spherically symmetric sector are given by
%Using the two-dimensional action \eqref{eq:2dtot}, the second junction condition for the theory \eqref{QTaction} on spherically symmetric metrics adopts the following form:
\begin{align}\label{2j}
\Pi_{AB}^--\Pi_{AB}^+=8\pi G_N S_{AB}\,, %&  \quad \Pi_{AB}=\Pi_{\tau \tau}+\frac{g_{ij}}{2} \Pi_{ij}\,, 
%\\ \Pi_{\tau\tau}=\frac{\varepsilon}{\varphi^{2}} F_3\,, & \quad g^{ij} \Pi_{ij}= -\frac{\varepsilon}{\varphi\dot \varphi} \frac{\mathrm{d}}{\mathrm{d} \tau} \left ( \varphi^{2} \Pi_{\tau \tau} \right)\,,
\end{align}
where %$i,j$ stand for those components along the angular directions of \eqref{Nf} and 
$ \Pi_{AB}^\pm $ stands for $\Pi_{AB}$ evaluated on each side of $\Sigma$.  The relevant components read
\begin{align}\label{piti}
    \Pi_{\tau \tau}&=\frac{2\varepsilon}{\varphi}\int_0^{n^{\mu}\partial_{\mu}\varphi}{\rm d}z\,h'\left(\frac{1+\varepsilon \dot\varphi^2-\varepsilon z^2}{\varphi^2}\right)\, , \\
   g^{ij} \Pi_{ij}&=-\frac{\varepsilon}{\varphi \dot\varphi}\frac{\diff}{\diff \tau}\left(\varphi^2 \Pi_{\tau\tau} \right)
\end{align}
where $i,j$ are indices on the angular coordinates. Hence, imposing Eq.\,(\ref{2j}) for the $\tau\tau$ component automatically enforces it for the rest of components.

\subsection{Oppenheimer-Snyder collapse}
Let us consider the standard case of Oppenheimer-Snyder collapse. For that, we take a perfect-fluid star made of pressureless matter, namely, we assume the matter stress tensor to be given by
\begin{equation}
T_{ab}=\rho(\tau) u_a u_b \,,
\end{equation}
where $\rho(\tau)$ is the fluid density, $u^a=\delta_\tau^a$ its $4$-velocity and $\tau$ the proper time of the fluid.

In the interior of the star, the metric is given by an FLRW spacetime of the form
\begin{equation}\label{ca}
\mathrm{d}s^2=-\mathrm{d}\tau^2+a(\tau)^2 \left[  \frac{\mathrm{d}\eta ^2}{1- \eta^2}+\eta^2\mathrm{d}\Omega_{2}^2\right]\, ,
\end{equation}
 where the scale factor satisfies the modified Friedmann equation \eqref{eq:cosmonpqt} with $w=0$,
% \begin{equation}\label{Frie}
%h(\Phi)=\frac{8\pi G}{3}\rho(\tau)\,  , \quad \text{where} \quad \Phi  \equiv \frac{1+ \dot a(\tau)^2}{a(\tau)^2} \, .
%\end{equation} 
%The continuity equation can be readily integrated, 
%\begin{equation}
%    \dot\rho(\tau)+\frac{3\rho(\tau)\dot a(\tau)}{a(\tau)}=0\,\, \Rightarrow \, \,\rho(\tau)=\frac{3\mathsf{a}}{8\pi G a(\tau)^3}\, ,
%\end{equation}
%for some constant $\mathsf{a}$, so that \req{Frie} becomes
 \begin{equation}\label{Friedd}
h(\Phi)=\frac{\mathsf{a}}{a(\tau)^3}\,  , %\quad \text{where} \quad \Phi  \equiv \frac{1+ \dot a(\tau)^2}{a(\tau)^2} \, ,
\end{equation} 
which is similar to the black hole equation \req{eomh}, as noted before. % if we perform the identifications $\Phi \leftrightarrow \psi$, $2GM \leftrightarrow \mathsf{a}$ and $r \leftrightarrow a(\tau)$.
On the other hand, as a consequence of Birkhoff's theorem, the exterior solution is given by
 \begin{equation}\label{f}
\mathrm{d}s^2=-f(r)\mathrm{d}t^2+\frac{\diff r^2}{f(r)}+r^2\diff \Omega_2^2\, ,
\end{equation}
where $f(r)$ is determined by \req{eomh}.  

\begin{figure}[t!]
\centering 
\includegraphics[width=0.47\textwidth]{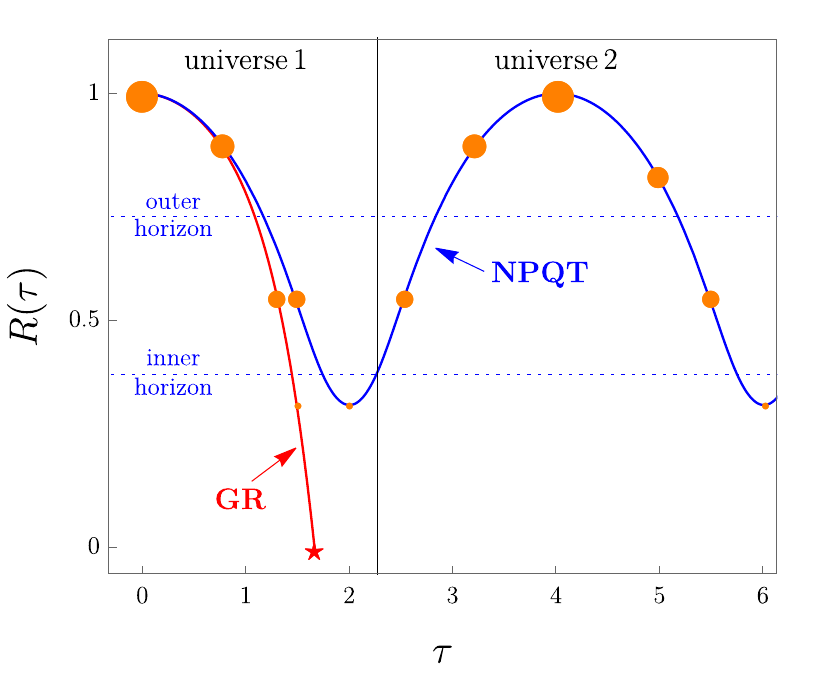}
\caption{We plot the coordinate radius (normalized by its initial value $R_0$) of a pressureless dust star as a function of its proper time as it undergoes gravitational collapse for Einstein gravity (red) and for a NPQT theory with gravitational couplings given by \req{coup} (blue). In GR, the star leaves behind a Schwarzschild black hole and reaches zero size (highlighted by a red star) after a finite proper time. On the other hand, for NPQT, the star reaches a minimum size inside the inner horizon of the regular black hole it creates, undergoing a bounce. It starts growing again and crosses the inner horizon and outer horizons of a white hole in a new universe, where it reaches its original size. The process is then restarted and repeated indefinitely. In the plot we have set $\rho_0=9/(32\pi G_{\rm N})$, $\alpha=1/8$.  }
\label{fig:modiH0}
\end{figure}

Let us parametrize the star surface, as seen from the inside, by $\eta=\eta_0$. Then, the induced metric reads
\begin{equation}
\diff s^2_-=-\mathrm{d}\tau^2+a(\tau)^2 \eta_0^2\mathrm{d}\Omega_{2}^2\, ,
\end{equation}
where $\eta_0$ is then star surface position in the comoving coordinates. In the exterior coordinates, we may parametrize the star surface by  $r=R(\tau)$, $t=T(\tau)$, and the corresponding induced metric reads in turn,
\begin{align}\notag
\diff s^2_+=&-\left(f(R(\tau)) \dot T(\tau)^2 - \frac{\dot R(\tau)^2}{f(R(\tau))}\right) \diff \tau^2\\ & + R(\tau)^2 \diff \Omega_{2}^2\, .
\end{align}
%outside, by $r=R(\tau)$, $t=T(\tau)$. Then, the induced metric reads
%The induced metric on the star surface as seen from the outside is given by
%where we parametrized the surface by $r=R(\tau)$, $t=T(\tau)$. 
%In the star interior coordinates, the surface can be parametrized by 
%From the interior perspective, the surface is parametrized by 
%$\eta=\eta_0$, so the induced metric reads in turn,
Now, the first junction condition requires that both induced metrics match. This yields
\begin{align}\label{jun1}
f(R(\tau))^2 \dot T(\tau)^2&=f(R(\tau))+\dot R(\tau)^2  \, , \\ R(\tau)^2&=a(\tau)^2 \eta_0^2\, . \label{ratu}
\end{align}
The second junction condition reads in turn,
%On the other hand, from the second junction condition we have
\begin{equation}
\Pi_{\tau\tau}^+ = \Pi_{\tau \tau}^- \, ,
\end{equation}
%in the case of a pressureless fluid
where we took into account that the stress tensor at the interface vanishes, $S_{AB}=0$. %, since no surface energy density on the interface is considered. 
Using \req{piti}, we have
\begin{equation}
\Pi_{\tau\tau}^{\pm}=\frac{2}{\varphi}\int_0^{n_{\pm}^\mu \partial_\mu \varphi_{\pm}} \diff z \, h'\left(\frac{1+\dot\varphi^2-z^2}{\varphi^2} \right)\, ,
\end{equation}
where the normal vectors read, respectively,
\begin{align}
n_+&=\frac{\dot R(\tau)}{f(R)} \partial_{t}+ f(R)\dot T(\tau) \partial_r\, ,\\ 
n_-&=\frac{\sqrt{1-\eta_0^2}}{a(\tau)}\partial_\eta \, ,
\end{align}
and $\varphi_+=r$, $\varphi_-=a(\tau) \eta$. From this, we find
\begin{align}
n_{+}^\mu \partial_\mu \varphi_+&=f(R(\tau))\dot T(\tau)\, ,\\
n_{-}^\mu \partial_\mu \varphi_-&=\sqrt{1-\eta_0^2}\, .
\end{align}
Hence, using the relation \req{ratu} between $R(\tau)$ and $a(\tau)$, the second junction condition simply reduces to
%\begin{equation}
%\frac{(D-2)}{R}\int_0^{f \dot T} \diff z \, h'\left(\frac{1+\dot R^2 -z^2}{R^2} \right) = \frac{(D-2)}{R}\int_0^{\sqrt{1-\eta_0^2}} \diff z \, h'\left(\frac{1+\dot R^2 -z^2}{R^2} \right) \, ,
%\end{equation}
%where we already imposed the relation between $R(\tau)$ and $a(\tau)$ coming from the first junction condition. This simply reduces to
\begin{equation}
f(R(\tau))\dot T(\tau)= \sqrt{1-\eta_0^2}\, .
\end{equation}
Combining this with Eq.\,\req{jun1} we find 
\begin{equation}\label{geodesic}
\dot R(\tau)^2 +\eta_0^2= 1 -f(R(\tau))\,  .
\end{equation}
This is nothing but the equation of  a timelike radial geodesic on the black hole background. %with an energy $E^2=1-\eta_0^2$. 
Hence, analogously to the higher-dimensional cases studied in \cite{Bueno:2025gjg}, for general four-dimensional NPQTs, each point in the star follows a geodesic. Without loss of generality, we can assume $\dot R(0)=0$, \ie we set $\tau=0$ as the moment in which the collapse starts. Then, the above equation reads
%that the star has not started collapsing yet at $\tau=0$, this can be alternatively written as
\begin{equation}\label{geodesic1}
\dot R(\tau)^2 +f(R(\tau)) =f(R_0)\,  ,
\end{equation}
where we called $R_0\equiv R(0)$ to the star initial coordinate radius. The ADM mass of the corresponding black hole or horizonless solution --- depending on the value of the mass --- is related with $R_0$ and the initial density of the star $\rho_0$ according to
\begin{equation}
  M=\frac{4\pi}{3} \rho_0 R_0^3\,,
\end{equation}
as one could have heuristically expected. Therefore, after fixing the theory, the gravitational collapse will be fully controlled by two physical parameters: $R_0$ and $\rho_0$. %This is illustrated in Figure \ref{fig:modiH2}.

As argued in \cite{Bueno:2025gjg}, the fact that the metric function behaves as $f(R)\simeq 1- R^2/C$ with $C>0$ near $R=0$ enforces the star to experience a bounce after reaching some minimum radius inside the inner horizon of the black hole it creates --- provided the original mass of the star spacetime is greater than $M_{\rm cr}$ --- through its collapse.

An explicit model is presented in Fig.\,\ref{fig:modiH0}, where the difference between the GR and NPQT behaviors is manifest. While in GR the star reaches zero size after a finite proper time, for the NPQT model the star never shrinks beyond certain radius. Once this minimum size is reached (something that always occurs inside the inner horizon of the black hole it leaves behind), the star experiences a bounce. It grows until it reaches its original radius in a new universe, where the process starts over again.

%Defining $x\equiv R(\tau)/R_0$ and
%\begin{equation}
%    \xi_0\equiv \frac{2G_{\rm N}M}{R_0^3}=\frac{8\pi G_{\rm N}}{3}\rho_0
%\end{equation}
%the star collapse equation reads in that case
%\begin{equation}
%    \dot x^2=\frac{x^2\xi_0}{\sqrt{x^6+\alpha^2\xi_0^2}}-\frac{\xi_0}{\sqrt{1+\alpha^2\xi_0^2}}
%\end{equation}
%where 

\subsection{Thin-shell collapse}
Consider now a thin shell of pressureless matter. In this case, the surface stress tensor is non-vanishing and reads $S_{AB} = \sigma u_A u_B$, where $\sigma$ is the surface energy density. The components of the surface stress tensor are then given by
%Let us now consider the collapse of a thin spherical shell of pressureless matter (``dust''). The surface stress-energy tensor takes the form $S_{AB} = \sigma u_A u_B$, where $\sigma$ is the surface energy density of the matter and $u_A$ is its $D$-velocity. In a proper time (denoted by $\tau$) parametrization of the shell, the components of the surface stress-energy tensor are simply
\be 
S_{\tau\tau} = \sigma \, , \quad S_{ij} = 0 \, .
\ee
%where $i,j$ are the angular components. 
For each $\tau$, we denote the shell radius by $R(\tau)$. Then, for $r < R(\tau)$ the metric is just Minkowski spacetime. On the other hand, by virtue of Birkhoff's theorem, the exterior of the shell, $r > R(\tau)$, is given by Eq.\,(\ref{f}), where $f(r)$ is determined by \req{eomh}. 
%At a given moment of proper time, we set the radius of the shell to $r = R(\tau)$. 
%Inside the shell, $r < R(\tau)$,  we take the metric to be Minkowski space. By Birkhoff's theorem, the exterior of the shell, $r > R(\tau)$, is necessarily the unique solution of~\eqref{eom}. 
Hence, the metric in the inside $(-)$ and the outside $(+)$ of the shell read, respectively, 
\be 
{\rm d} s_\pm^2 = - f_\pm (r)  {\rm d}t_\pm^2 + \frac{ {\rm d}r ^2}{f_\pm(r)} + r^2 {\rm d} \Omega^2_{2} \, ,
\ee
where $f_-(r) = 1$ and $f_+(r)$ is the solution to~\eqref{eomh}.

\begin{figure}[t!]
\centering 
\includegraphics[width=0.47\textwidth]{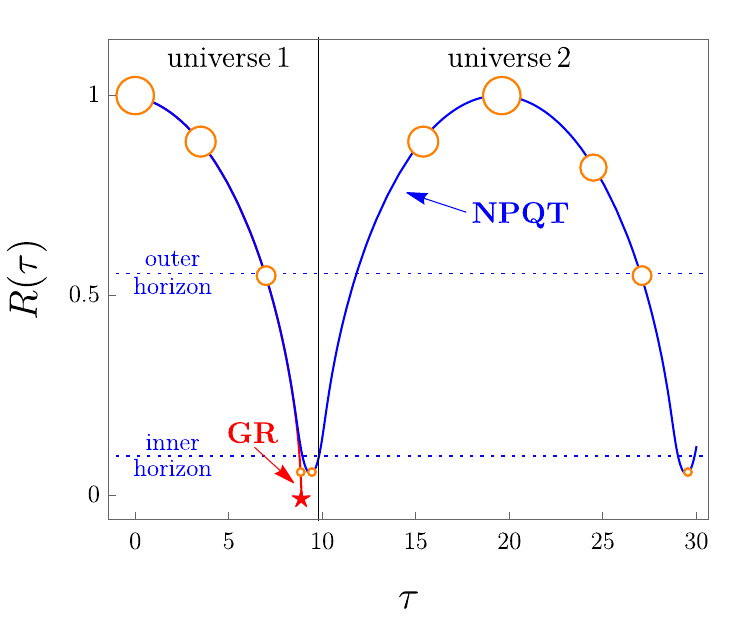}
\caption{We plot the coordinate radius (normalized by its initial value, $R_0$) of a spherical thin shell as a function of its proper time as it undergoes gravitational collapse for Einstein gravity (red) and for a NPQT theory with gravitational couplings given by \req{coup} (blue). The behavior is completely analogous to the one explained in the caption of Fig.\,\ref{fig:modiH0} for the collapse of a pressureless dust star. In the plot we have set $G_{\rm N}M=1$, $G_{\rm N}m=6/5$, $\alpha=1/8$.   }
\label{fig:modiH2}
\end{figure}

Just like in the dust star case, the two charts must be joined at the shell location using the modified Israel junction conditions. These determine the shell trajectory, which we parametrize as $(t_\pm, r) = \left(T_\pm(\tau), R(\tau) \right)$. The induced metric on the shell reads then
\be 
{\rm d}s_\Sigma^2 = - \left(f_\pm(R) \dot{T}_\pm^2 - \frac{\dot{R}^2}{f_\pm(R)} \right) \diff \tau^2 + R(\tau)^2 {\rm d} \Omega_{2}^2 \, ,
\ee
which also equals
\be 
{\rm d}s_\Sigma^2 = -  \diff \tau^2 + R(\tau)^2 {\rm d} \Omega_{2}^2 \, ,
\ee
since $\tau$ is the shell proper time. Taking this into account, the first junction condition imposes\footnote{\label{fn:signs} For the variables affected by the ``$+$'' subindex, condition \eqref{eq:juncon} must be accompanied by a global $\pm$ to account for the fact that the left-hand side may change its sign in the black hole interior \cite{Bueno:2024zsx}. In such a case, Eq.\,\eqref{sdd} must also be slightly modified.}
%and demanding continuity (taking into account that $\tau$ is the proper time), we find that
\be 
f_{\pm}(R)\dot{T}_\pm = \sqrt{f_\pm(R) + \dot{R}^2} \equiv \beta_\pm \,  .
\label{eq:juncon}
\ee
On the other hand, the relevant component of the boundary equation follows from Eq.\,\req{piti} and reads
\be 
\Pi_{\tau\tau}^\pm = \frac{2}{R} \int_0^{\beta_\pm} {\rm d}z \, h' \left(\frac{1 + \dot{R}^2 - z^2}{R^2} \right) \, .
\ee
The second junction condition imposes
%and we need to impose
\begin{align}
\Pi_{\tau\tau}^- - \Pi_{\tau\tau}^+ &= 8 \pi G_N \sigma \, ,
\\
\frac{{\rm d}}{{\rm d} \tau} \left[R^{2}\left(\Pi_{\tau\tau}^- - \Pi_{\tau\tau}^+ \right) \right] &= 0 \, .
\end{align}
%where 
%\be 
%\Pi_{\tau\tau}^\pm = \frac{(D-2)}{R} \int_0^{\beta_\pm} {\rm d}z  h' \left(\frac{1 + \dot{R}^2 - z^2}{R^2} \right) \, .
%\ee
The second equation implies the conservation of the shell's proper mass, 
\be 
m \equiv 4\pi R(\tau)^{2}  \sigma(\tau)   = {\rm constant} \, .
\ee
The first can be written as an integro-differential equation for $R(\tau)$, namely,
\begin{equation}\label{sdd}
\frac{m}{3R^{3}}=\int_{R}^{\infty} \frac{\mathrm{d}r \,M }{r^4 \sqrt{1+\dot R^2-\frac{R^2}{r^2}\left[1-f(r) \right]}}\, .
\end{equation}
This equation can be recast as
\begin{equation}\label{jijo}
    \dot R^2+V(R)=\frac{M^2}{m^2}-1\, ,
\end{equation}
for some theory-dependent effective potential. In the case of Einstein gravity, the potential reads
\begin{equation}
    V(R)=-\frac{G_N M}{R}-\frac{G_N m}{4R^2}\,, 
\end{equation}
which is monotonically decreasing as a function of $R$. Hence, in that case any shell starting with a finite radius $R(0)\equiv R_0$ reaches zero size after a finite proper time. On the other hand, for non-polynomial QT theories with regular black holes, the potential behaves as
\begin{equation}
    V(R)=-\frac{R^2}{C} \quad {\rm as}\quad R\rightarrow 0\, ,
\end{equation}
which means that it possesses a minimum at some intermediate theory-dependent value of $R$. Hence, in this case the shell experiences a bounce after reaching a minimum size, analogously to the dust star considered in the previous subsection. 

We have performed the explicit integration of Eq.\,\req{jijo} in the case of the non-polynomial QT corresponding to the gravitational couplings given by \req{coup} for a particular choice of parameters. The result is shown in Fig.\,\ref{fig:modiH2}, where it is clear that the evolution of the shell is very similar to the one of the dust star considered in the previous subsection. Assuming that $M$ is sufficiently large to produce a black hole, the shell will collapse until it reaches a minimum coordinate radius inside the inner horizon of the black hole it leaves behind. It bounces back and appears in a new universe through a white hole, where it eventually reaches its initial size. This process repeats endlessly.

%A shell with proper mass greater than $M$ collapses until it reaches a minimum coordinate radius inside the inner horizon of the black hole it leaves behind. It bounces back and appears in a new universe through a white hole, where it eventually reaches its initial size. This process repeats endlessly.

\section{Discussion}

We have presented a set of four-dimensional non-polynomial gravitational theories that have the following properties:
\begin{itemize}
    \item In spherical symmetry, the equations of motion are second-order and match a naive four-dimensional limit of higher-dimensional QT gravities. 
    \item The theories have a Birkhoff theorem, ensuring the general spherical solution is static and characterized by just the mass. 
    \item Linear perturbations on top of maximally symmetric backgrounds obey second-order equations.
    \item Upon including an infinite tower of these non-polynomial terms, black hole and cosmological singularities are eliminated, yielding explicit regular black hole solutions and bouncing cosmologies. \item In spherical symmetry, the dimensional reduction of the theories produces a particular family of two-dimensional Horndeski gravities. Based on this, we studied nonsingular gravitational collapse, showing that collapsing dust and thin shells produce the RBH solutions of the theories under consideration.
\end{itemize}
All together, these results comprise the natural four-dimensional extension of the program of work recently undertaken in five and higher dimensions~\cite{Bueno:2024dgm, Bueno:2024zsx, Bueno:2024eig, Bueno:2025gjg}. While some of the details presented here have been previously known, our work represents (to the best of our knowledge) the most complete study of the existence and formation of RBHs in four-dimensional gravity. Our setup can be used to study questions about RBH dynamics in four dimensions, and may therefore have direct astrophysical relevance. Let us now discuss some of the most important structural considerations for future work.

The set of non-polynomial theories we presented were chosen for their particular properties in spherical symmetry. In the full space of non-polynomial actions there will certainly exist \textit{degeneracies}. That is, we expect the existence of a potentially large class of theories which when restricted to spherical symmetry all produce the same reduced action and solutions, an idea which is further reinforced by the works~\cite{Colleaux:2017ibe,Colleaux:2019ckh}. Investigations that probe beyond the spherical sector (such as perturbations) will be sensitive to the differences between members of these equivalence classes of theories. Hence, an important prerequisite for any such study is performing a detailed analysis of these degeneracies. 

In another vein, the specific non-polynomial actions we analyzed were chosen because they agree with a four-dimensional limit of higher-dimensional quasi-topological gravities; as a result, when resummed they satisfy Markov’s limiting-curvature hypothesis and retain several other desirable features. However, we expect the existence of non-polynomial theories that yield second-order equations of motion in spherical symmetry yet are inequivalent to those presented above. Specifically, one could construct non-polynomial theories whose dimensional reduction produces more general two-dimensional Horndeski models~\cite{Kunstatter:2015vxa, Carballo-Rubio:2025ntd}, which could be specially relevant beyond spherical symmetry~\cite{KolarPrep}. Systematically charting this space of theories is an interesting direction for future work.

% Beyond degeneracies, we expect the existence of non-polynomial theories which have second-order equations of motion in spherical symmetry yet differ from those we have presented here. The non-polynomial actions we have selected are chosen to match with a four-dimensional limit of higher-dimensional QTs.  As such they inherit a number of desirable properties, such as satisfying Markov's limiting curvature hypothesis when resummed. One can easily imagine non-polynomial theories which produce more general 2-dimensional Horndeski theories upon dimensional reduction. Beyond this, non-polynomial QTs will also exist for metrics outside of spherical symmetry~\cite{KolarPrep}. Understanding the space of these theories is an interesting and important line of investigation. 

A potentially serious concern is whether non-polynomial actions can be dynamically sensible or whether the action itself may become highly singular. For example, the representative actions we have presented feature denominators which could in principle vanish. While we expect singular behaviour to arise at least for \textit{some} non-polynomial theories, it is not clear whether it will happen for \textit{all} such theories --- and, ultimately, it is only necessary that \textit{one} representative resummation be sensible. We do not have an answer to this question at this stage, but let us make a positive remark based on analogy with much simpler systems. The action governing the motion of a point particle in special relativity is of a nonpolynomial nature, $\mathcal{L}_{\rm pp} = \sqrt{1 - \dot{x}^2/c^2}$, and would become pathological if $\dot{x} > c$. However,  this is avoided as any motion that starts with $\dot{x} < c$ must remain in this sector. Hence, one may hope that there exist non-polynomial actions which dynamically enforce their own consistency. 

The strongest motivation for the higher-dimensional work in~\cite{Bueno:2024dgm} is the connection with effective field theory. In the perturbative regime, any polynomial action (and certain terms involving covariant derivatives) can be mapped via field redefinition to a QT gravity.\footnote{It is expected that any (vacuum) effective action can be captured by a suitable QT theory. However, even if that were not the case, terms with covariant derivatives are unlikely to spoil the regular black hole solutions, as such terms necessarily vanish in the de Sitter regions found in the RBH core.} Hence, even though QT theories are engineered to have specific properties, they are not fine-tuned in the space of all theories and therefore particularly well-motivated. It is not clear if similar motivation exists for non-polynomial actions, and this must be addressed. If motivation cannot be obtained on broad principles like those for polynomial QTs, perhaps more general criteria such as limiting curvature, stability, or sensible dynamics can be used to motivate particular instances of non-polynomial actions, such as in~\cite{Yoshida:2017swb}. 

Ultimately, the question that~\cite{Bueno:2024dgm} motivates is whether resumming four-dimensional gravitational effective theories suffices to resolve singularities. Our results show that there exist covariant, four-dimensional theories that admit regular black holes as unique solutions and for which collapse dynamics is tractable in the spherical sector. How this connects to the broader question remains to be seen. Either we will find that, as in higher dimensions, the resummation of these terms can resolve at least some of the singularities of Einstein gravity, or we will learn that our four-dimensional world is even more special than we thought.

%is whether resumming classical corrections to four-dimensional general relativity suffices to resolve singularities. Our results show that there exist four-dimensional theories that admit regular black holes as unique solutions and for which collapse dynamics is tractable in the spherical sector. How this connects to the broader question remains to be seen. Either we will find that, as in higher dimensions, the resummation of these terms can resolve at least some of the singularities of Einstein gravity, or we will learn that our four-dimensional world is even more special than we thought.

\section*{Acknowledgements}

We thank Ivan Kol{\'a}{\v r} for useful discussions on the upcoming related work~\cite{KolarPrep}. PB was supported by a Ram\'on y Cajal fellowship (RYC2020-028756-I), by a Proyecto de Consolidación Investigadora (CNS 2023-143822) from Spain’s Ministry of Science, Innovation and Universities, and by the grant PID2022-136224NB-C22, funded by MCIN/AEI/ 10.13039/501100011033/FEDER, UE.
The work of PAC was supported by a fellowship from ``la Caixa'' Foundation (ID 100010434) with code LCF/BQ/PI23/11970032 and by a Ram\'on y Cajal fellowship (RYC2023-044375-I) from Spain’s Ministry of Science, Innovation and Universities.  \'AJM  was supported by a Juan de la Cierva contract (JDC2023-050770-I) from Spain’s Ministry of Science, Innovation and Universities.

\bibliographystyle{JHEP-2}
\bibliography{Gravities}
%\vspace{1cm}
\noindent 

\end{document}